\documentclass[
	 aps, prx
	,superscriptaddress
	,reprint 
	,longbibliography
	]{revtex4-2}

\usepackage{siunitx} 

\usepackage{amsmath} 
\usepackage{amssymb}
\renewcommand*{\vec}[1]{\mathbf{#1}}

\usepackage{graphicx} 

\usepackage[utf8]{inputenc}
\usepackage[T1]{fontenc}

\usepackage{cancel}

\usepackage[table,xcdraw]{xcolor}

\newcommand{\univieA}{Faculty of Physics, University of Vienna, 1090 Vienna, Austria}
\newcommand{\univieB}{Research Platform MMM Mathematics - Magnetism - Materials, University of Vienna, Vienna 1090, Austria}
\newcommand{\PSIMesosys}{Laboratory for Multiscale Materials Experiments, Paul Scherrer Institute, 5232 Villigen PSI,~Switzerland}
\newcommand{\ETHMesosys}{Laboratory for Mesoscopic Systems, Department of Materials, ETH Zurich, 8093 Zurich,~Switzerland}
\newcommand{\nanogune}{CIC nanoGUNE BRTA, 20018 Donostia-San Sebasti\'{a}n, Spain}
\newcommand{\ikerbasque}{IKERBASQUE, Basque Foundation for Science, 48013 Bilbao, Spain}
\newcommand{\stockholm}{Department of Physics, Stockholm University, 106 91 Stockholm, Sweden}

\begin{document}
	

\title[ %
	Chiral switching in artificial square ice
	]{ %
	Chiral switching and dynamic barrier reductions in artificial square ice
	}

\author{Na{\"e}mi~Leo}
\email{n.leo@nanogune.eu}
\affiliation{\nanogune}

\author{Matteo~Pancaldi}
\affiliation{\stockholm}

\author{Sabri~Koraltan}
\affiliation{\univieA}

\author{Pedro~Villalba~Gonz{\'a}lez}
\affiliation{\nanogune}

\author{Claas~Abert}
\affiliation{\univieA}
\affiliation{\univieB}

\author{Christoph~Vogler}
\author{Florian~Slanovc}
\author{Florian~Bruckner}
\author{Paul~Heistracher}
\affiliation{\univieA}

\author{Kevin~Hofhuis}
\affiliation{\ETHMesosys}
\affiliation{\PSIMesosys}

\author{Matteo~Menniti}
\affiliation{\nanogune}

\author{Dieter~Suess}
\affiliation{\univieA}
\affiliation{\univieB}

\author{Paolo~Vavassori}
\affiliation{\nanogune}
\affiliation{\ikerbasque}

\begin{abstract}
	
	
	Collective dynamics in lithographically-defined artificial spin ices offer profound insights into emergent correlations and phase transitions of geometrically-frustrated Ising spin systems.
	The understanding of experimentally-observed temporal evolution of extended spin ices are often guided and supported by model predictions, for example from kinetic Monte-Carlo simulations. 
	This coarse-grained approach, which disregards microscopic details of the moment reversal, allows to simulate systems with a large number of moments evolving over long time scales, which otherwise would be too computationally-costly to be implemented in full micromagnetic simulations.
	To obtain correct relaxation time scales and spatial correlations, kinetic Monte Carlo simulations rely instead on the precise knowledge of the rates for individual moment reversal.
	These rates are determined by the switching barriers which, in many cases, are derived from simplified or approximative assumptions only, which do not take into account the full physical picture of nanomagnetic switching.  
	
	In this work, we describe how the immediate magnetic environment of a nanomagnet reversing via quasi-coherent rotation can induce clockwise and counter-clockwise switching channels with different barrier energy. 
	We compare predictions from a perturbative model to switching barriers obtained from micromagnetic string-method simulations for two different -- exchange- vs.\ magnetostatically-dominated -- artificial square ice geometries. 
	Taking into account the spatial extension and non-uniform magnetic behaviour, we find further reductions and enhanced barrier splitting, especially in the case of magnetostatically-dominated nanomagnets.
	These modifications of the switching barriers lead to exponentially enhanced relaxation kinetics, especially in the limit of rare events. From kinetic Monte-Carlo simulations we find that the evolution invoking split barriers yields much faster relaxation time scales and results in different spatial moment configurations compared to the often-employed mean-field transition barriers. 
	Our results highlight how the local magnetic environment can significantly enhance the transition kinetics and affect emergent correlations, even without invoking defects or additional anisotropies.
	These findings are thus of integral importance to achieve realistic kinetic Monte Carlo simulations of emergent correlations in extended artificial spin systems, magnonic crystals, or the evolution of small-scale nanomagnetic logic circuits. 
	
\end{abstract}



\maketitle


\begin{figure*}[tb]
	\centering
	\includegraphics[width = 176mm]{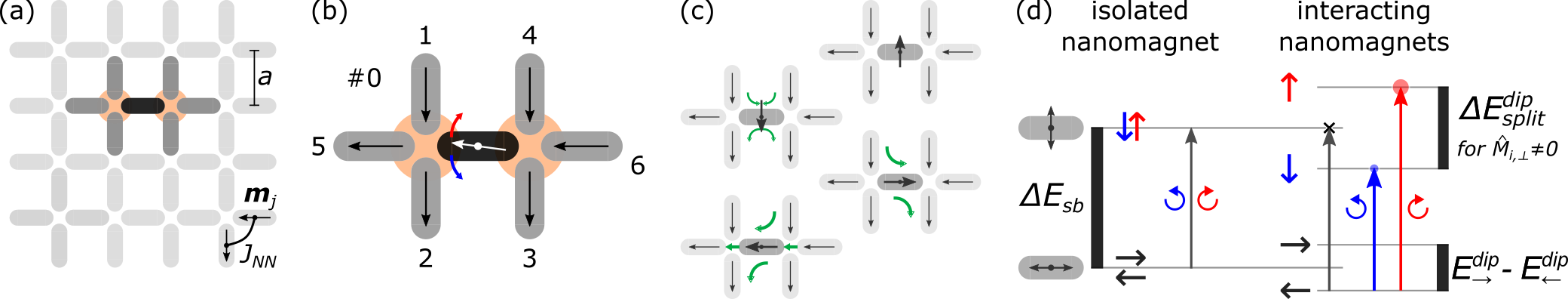}
	\caption{%
		\textbf{Switching barriers in artificial square ice.}
		(a)~Nanomagnets are arranged on a square lattice with periodicity $a$. The strongest mutual interaction $J_\textit{NN}$ acts between perpendicular nearest-neighbour moments $\vec{m}_j$. 
		(b)~Fully-magnetised double-vertex state $\#0$, with enumeration of moments. The central nanomagnet (black) can rotate either clockwise (red arrow) or counter-clockwise (blue arrow) from left ($\leftarrow$) to right ($\rightarrow$).
		(c)~Energetically-favourable states, on the left, feature more head-to-tail configurations between the central and neighbouring moments (green arrows).
		(d)~In a pertubative picture, the switching barrier energies can be obtained by adding the interaction energies to the switching barrier ${\Delta E_\text{sb}}$ of an isolated nanomagnet.
		If only the energies of the equilibrium configurations ($\leftarrow$, $\rightarrow$) are taken into account, a mean-field barrier is obtained (gray arrow to cross).
		In case the environment features a perpendicular magnetisation, i.e.\ $\hat{M}_{i,\perp}\ne0$, the high-energy states ($\uparrow$, $\downarrow$) will split, and thus yields separate transition barriers for clockwise (red) and counter-clockwise (blue) rotation.	
	}
	\label{fig:square_ice-overview}
\end{figure*}


Artificial spin ice systems are lithographically-created lattices of elongated single-domain nanomagnets, and have been designed to investigate the effect of correlations and the onset of long-range order in frustrated two-dimensional magnetic lattices \cite{2006Wang,2013Nisoli,2013Heyderman,2019Skjaervoe}.

Of particular interests is the evolution of extended spin ice lattices from a field-saturated state towards an energetically favourable (ground) state, driven by thermally-activated reversal of individual nanomagnets. Such experiments have been performed mainly using photoelectron emission microscopy, and gave valuable insight on the relaxation process and the formation of spatial correlations \cite{2012Porro,2013Zhang,2013Farhan,2013Farhan_a,2014Kapaklis}.
%
%
These results are often compared to model predictions for the temporal and spatial evolution from kinetic Monte Carlo (kMC) simulations. Their major advantage over full micromagnetic simulations is that they are less computational costly, and thus can be extended to larger systems and longer time scales. 
%
%
To match the measured experimental time scales, however, the two main parameters determining the switching barriers used in the kMC simulations -- the single-particle barrier and the interaction strength -- are often adjusted \cite{2013Farhan,2014Kapaklis,2016Andersson,2017Morley}. These changes, however, are usually only loosely justified by physical reasoning, and seldomly put onto consistent grounds.
Furthermore, an often-used mean-field approach does not take into account the freedom for clockwise or counter-clockwise rotations \cite{2013Farhan,2013Farhan_a,2014Thonig,2019Arava,2020Jensen_a}, which can lead to distinct switching barriers, as we showed previously \cite{2020Koraltan}. 

In this work, we derive that a net perpendicular field from a defect-free double-vertex environment acting on the switching nanomagnet enables favourable chiral reversal pathways in artificial square ice.
We compare switching barriers obtained from micromagnetic string-method simulations for exchange- and magentostatic-dominated geometries to those derived from simplified point-dipole predictions. We find that the latter consistently overestimates the barriers and underestimates the chiral splitting of the former, and are not applicable even with renormalised parameters in the case where non-coherent reversal modes are possible.

Reductions and splitting of the switching barriers lead to exponentially enhanced transition rates especially in the limit of rare events, as we show with a modified Arrhenius law. 
Using the rates for the chiral transition channels as input for kMC simulations, we find that the evolution of an extended square ice proceeds much faster, and involves different spatial correlations when compared to a mean-field model.
The influence of the immediate environment on the nanomagnetic switching thus is a key ingredient to correctly model the relaxation dynamics of artificial spin ices, as well as of functional magnonic materials and small-scale circuits for computation. We therefore expect our results to be relevant to different communities making use of thermally-driven relaxation of interacting nanomagnets.


This work is structured into three sections: 
In Sec.~\ref{sec:switching_theory} a basic understanding is derived on how the magnetic environment can lead to chiral switching channels in artificial square ice.
Sec.~\ref{sec:simulations} compares point-dipole model predictions to micromagnetic simulations of nanomagnets of different dimensions and material parameters. 
In Sec.~\ref{sec:kinetics}, ramifications of the modified switching  barriers on the switching rates of single nanomagnets and relaxation kinetics of extended artificial square ice are discussed.

\section{Chiral moment reversal} 
\label{sec:switching_theory}

\begin{figure}[bt]
	\centering
	\includegraphics[width=77.909mm]{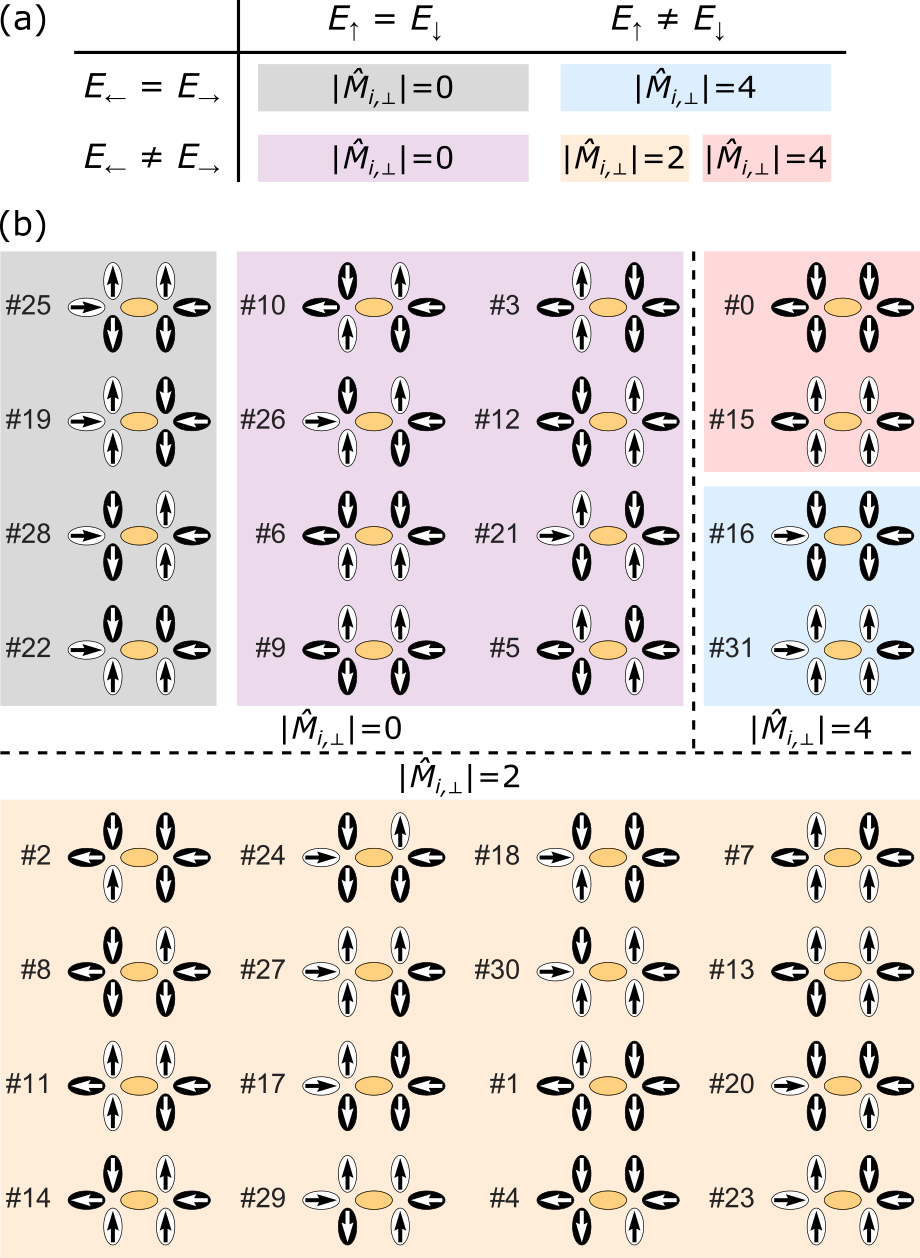}
	\caption{%
		\textbf{Enumeration of double-vertex states.} 
		(a)~Based on the dipolar energies $E_{i,k}^\mathrm{dip}$ of the environment configurations $\#i$ where the central moment can point in different directions, $k=\leftarrow$, $\rightarrow$, $\uparrow$, $\downarrow$, five classes can be distinguished (highlighted with different colours). Configurations with a magnetisation ${\hat{M}_{i,\perp}\ne0}$ perpendicular to the central nanomagnet feature distinct barriers for nanomagnet reversal via clockwise and counter-clockwise rotation.
		(b)~Environment configurations \#0 to \#31 sorted into the five categories.
		In total, 40 out of the 64 environment states promote switching with a favoured chirality.
		}
	\label{sfig:enumerate_of_states}
\end{figure}

\begin{figure*}[t!]
	\centering
	\includegraphics[width=164.813mm]{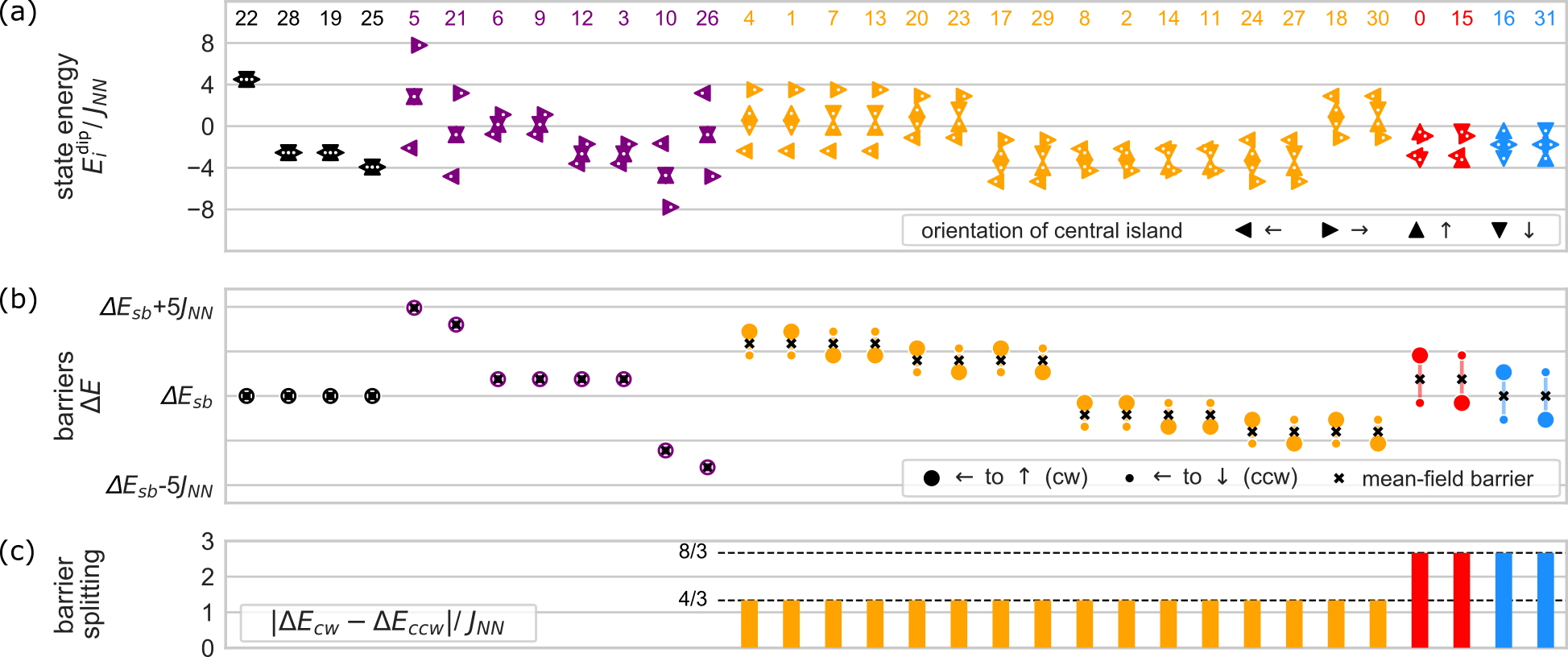}
	\caption{%
		\textbf{Barrier splitting in a double-vertex environment}, based on the point-dipole model for moment reversal.
		(a)~Normalised dipolar configuration energies $E_{i,k}/J_\mathrm{NN}^\mathrm{dip}$ of a central moment $k$ pointing to the left $k=\leftarrow$, top $k=\uparrow$, right $k=\rightarrow$, or bottom $k=\downarrow$ (as indicated with triangles) embedded in environments $\#i$ (labelled on the top). 
		The colour code corresponds to the scheme presented in Fig.~\ref{sfig:enumerate_of_states}.
		(b)~Chiral switching barriers ${\Delta E_{i,\leftarrow,\text{cw}}^\mathrm{dip}}$ and ${\Delta E_{i,\leftarrow,\text{ccw}}^\mathrm{dip}}$ are marked by large and small circles, respectively, and the mean-field barrier by a cross. 
		(c)~For environments with finite perpendicular magnetisation ${\hat{M}_{i,\perp}\ne 0}$ acting on the the central nanomagnet a splitting between clockwise and counter-clockwise barriers is observed (marked in yellow, red and blue). 
		The energy splitting predicted by Eq.~(\ref{eq:point_dipole_barrier_torque}), and normalised to $J_\textit{NN}^\mathrm{dip}$, is marked by dashed lines.
		}
	\label{sfig:point-dipole-splitting}
\end{figure*}

Artificial square ice is a lithographically-designed magnetic metamaterial with identical nanomagnets arranged on a square lattice with periodicity $a$, see Fig.~\ref{fig:square_ice-overview}(a). 

Due to their shape anisotropy, each stadium-shaped nanomagnet with length $l$, width $w$, and thickness $t$ is quasi-uniformly magnetised, and thus behaves like an Ising macrospin. 
Without applied field or interactions with neighbours, the magnetic moment will align with the long axis, i.e. to the left ($\leftarrow$) or right ($\rightarrow$). To spontaneous switch between these energetically-degenerate configurations (without loss of generality from $\leftarrow$ to $\rightarrow$), the moment rotates to overcome a metastable state for which the net moment points along the nanomagnets' short axis, i.e.\ $\uparrow$ or $\downarrow$.
The difference between the metastable and equilibrium micromagnetic energies $E^\mathrm{mm}$ gives the single-island switching barrier $\Delta E_\text{sb}$:
\begin{equation}
	\Delta E_\text{sb} 
	= E_{\updownarrow}^\mathrm{mm} - E_{\leftrightarrow}^\mathrm{mm} \, .
	\label{eq:single-moment-barrier}
\end{equation}
The value of ${\Delta E_\text{sb}}$ depends on the size, shape and material of the individual elements \cite{2000Cowburn,2001Usov,2013Chopdekar,2015Flovik}. 
For the assumption of uniform magnetisation, the barrier is given by the shape anisotropy ${\Delta E_\text{sb}^\text{shape}=K^\text{shape}V}$, with $V$ being the volume of the nanomagnet. Values of $K^\text{shape}$ are either tabulated \cite{1945Osborn} or can be calcuated via magnetostatic simulations \cite{oommf_userguide,2014Vansteenkiste}.

The mutual coupling between nanomagnets is given by magnetostatic interactions, which takes the following form between point-dipole moments $\vec{m}_i$ and $\vec{m}_j$ separated by a distance vector $\vec{r}_{ij}=\vec{r}_i - \vec{r}_j$: 
	\begin{eqnarray}
		E^{\mathrm{dip}} = 
			\frac{\mu_0}{4 \pi |\vec{r}_{ij}|^3} \left[ 
			\vec{m}_i \cdot \vec{m}_j 
			- 3\frac{\left( \vec{m}_i\cdot\vec{r}_{ij}  \right) \left( \vec{m}_j\cdot \vec{r}_{ij} \right)}{|\vec{r}_{ij}|^2}
			\right] 
			 . 
		\label{eq:energy-point-dipole}
	\end{eqnarray}
In artificial square ice the strongest coupling, denoted by $J_\textit{NN}$, occurs between nearest-neighbour nanomagnets at a \ang{90} angle, see Fig.~\ref{fig:square_ice-overview}(a). Using the lattice periodicity $a$ and the net moment ${m=|\vec{m}|=M_\text{sat}V}$ of a nanomagnet with volume $V$ and material saturation magnetisation $M_\text{sat}$, we define a convenient energy scale $J_\textit{NN}^\mathrm{dip}$:
	\begin{equation}
		J_\textit{NN}^\mathrm{dip} = \frac{3}{2\sqrt{2}} \frac{\mu_0}{\pi} \frac{m^2}{a^3} \, .
		\label{eq:JNN}
	\end{equation}

Due to the pronounced distance dependence, ${E^{\mathrm{dip}}\propto r^{-3}}$, the coupling between nanomagnets meeting at the vertex points highlighted in Fig.~\ref{fig:square_ice-overview}(a), is dominant over further-range interactions \cite{2002Politi,2018Macedo}.
Therefore, we investigate the switching barriers for moment reversal of a central nanomagnet under the influence of its closest neighbours only.

\subsection{Switching environments}
\label{sec:switching_environments}

For an infinite artificial square ice the environment that influences the reversal of a nanomagnet forms a double-vertex configuration, as depicted in Fig.~\ref{fig:square_ice-overview}(b). Here, each tip of the central nanomagnet (black) is in close interaction to three other nanomagnet (gray), whose magnetisation remains largely unchanged during the reversal of the central nanomagnet. The extended square lattice then can be considered as an infinite tiling of these motifs.

We denote each magnetic equilibrium configuration with a state number $\#i$ determined by the arrangement of the surrounding nanomagnets, and the orientation of the central (switching) nanomagnet, which can point to the left ($\leftarrow$) or to the right ($\rightarrow$). 
The state number $\#i$ can be obtained from the binary representation of the relative configuration of the six surrounding nanomagnet numbered $1$ to $6$ according to the scheme shown in Fig.~\ref{fig:square_ice-overview}(b):
\begin{eqnarray}
	i & = & \Sigma_{j=1}^{6} \, b_j \, 2^{j-1} \label{eq:state_number} \\
	\mathrm{with }\; b_{j} & = & 
		\begin{cases}
			0 & \text{if moment points down or to left}\\
			1 & \text{if moment points up or to right}
		\end{cases} \nonumber
\end{eqnarray} 
Half of the $2^{6}=64$ environment configurations of the double vertex are depicted in Fig.~\ref{sfig:enumerate_of_states}. The remaining states $\#i$ can be derived by applying a time reversal operation on the configurations ${\#(2^6-1-i)}$.

\subsection{Switching barriers from a point-dipole model}
\label{sec:point-dipole-barriers}

%
In the following, we discuss a perturbative approach to switching barriers in artificial square ice. Here, the single-nanomagnet barrier $\Delta E_\text{sb}$, Fig.~\ref{fig:square_ice-overview}(d), is modified due to energy contributions arising from the interactions with the immediate magnetic environment.

%
To discuss a specific example, we focus on the fully-magnetised double-vertex environment $\#0$ depicted in Fig.~\ref{fig:square_ice-overview}(b). The energy of the configuration where the central moment points (exactly) to the left, $\leftarrow$, is lower than when it points to the right, $\rightarrow$, where three magnetic charges meet at each vertex point. 
Using simplified assumptions and symmetry arguments, one can derive a mean-field switching barrier ${\langle\Delta E_{i}\rangle_{\leftarrow\text{ to }\rightarrow}^\mathrm{dip}}$ (as the average of the barriers for clockwise and counter-clockwise rotation, see Appendix~A in Ref.~\cite{2020Koraltan}). Its value depends solely on the energies of the equilibrium configurations, as indicated by the gray arrow in Fig.~\ref{fig:square_ice-overview}(e):
	\begin{equation}
		\langle\Delta E_{i}\rangle_{\leftarrow\text{ to }\rightarrow}^\mathrm{dip}
			= \Delta E_\text{sb}^\text{shape} + \frac{1}{2} \left(
				E_{i,\rightarrow}^\mathrm{dip} - E_{i,\leftarrow}^\mathrm{dip}
			\right) 
		\label{eq:mean-field-barrier}
	\end{equation}

%
This mean-field barrier, however, is missing a crucial point, as independent relaxation pathways via clockwise and counter-clockwise rotation of the central nanomagnet need to be considered, i.e.\ 
\begin{eqnarray}
\Delta E_{i, \leftarrow, \mathrm{cw}}^\mathrm{dip} 
& = & \Delta E_\mathrm{sb}^\text{shape} + \left( E_{i,\uparrow}^\mathrm{dip} - E_{i,\leftarrow}^\mathrm{dip} \right)	 \,	, 
\label{eq:chiral_barriers_cw}
\\
\Delta E_{i, \leftarrow, \mathrm{ccw}}^\mathrm{dip} 
& = & \Delta E_\mathrm{sb}^\text{shape} + \left( E_{i,\downarrow}^\mathrm{dip} - E_{i,\leftarrow}^\mathrm{dip} \right)	 \,	. 
\label{eq:chiral_barriers_ccw}
\end{eqnarray}
These barriers are not necessarily equivalent, as shown in Fig.~\ref{fig:square_ice-overview}(c): Due to their staggered spatial arrangement, the central moment in a fully magnetised environment \#0 will preferably align ferromagnetically with its neighbours, thus forming a head-to-tail flux-closure configuration which reduces the dipolar energy term in Eq.~(\ref{eq:energy-point-dipole}). 
Therefore, transitions via counter-clockwise rotations (blue) of the central nanomagnet will be largely favoured over those via clockwise rotations (red). 

From the dipolar energies for all environmental states and orientation of the central moment, Fig.~\ref{sfig:point-dipole-splitting}(a), one can obtain the respective clockwise and counter-clockwise switching barriers, Fig.~\ref{sfig:point-dipole-splitting}(b).
Under the assumption that $\uparrow$ and $\downarrow$ align perfectly along the short axis of the nanomagnets, the energy splitting between the configurations is symmetric around the mean-field barrier ${\langle\Delta E_{i}\rangle_{\leftarrow\text{ to }\rightarrow}^\mathrm{dip}}$ (marked by crosses), and equals  the three distinct values shown in Fig.~\ref{sfig:point-dipole-splitting}(c).

Barrier splitting occurs for all environments which feature a perpendicular effective field $\hat{M}_{i,\perp}$  generated by the neighbouring nanomagnets, that acts on the central nanomagnet [the indices $b_j$ are as defined in Eq.~(\ref{eq:state_number})]:
\begin{equation}
	\hat{M}_{i,\perp} = \Sigma_{j=1}^{4} \frac{\vec{m}_j}{|\vec{m}_j|} = \Sigma_{j=1}^{4} (-1)^{b_j} \, .
	\label{eq:Mperp}
\end{equation}
The normalised perpendicular magnetisation $\hat{M}_{i,\perp}$ can take the values of $0$ (black and purple in Figs.~\ref{sfig:enumerate_of_states} and \ref{sfig:point-dipole-splitting}), $\pm 2$ (orange), and $\pm 4$ (red and blue) only.
Thus, we can modify Eq.~(\ref{eq:mean-field-barrier}) to include an additional energy term:
	\begin{equation}
		\Delta E_{i,\leftarrow, \mathrm{cw/ccw}}^\mathrm{dip}
		=
		\langle\Delta E_{i}\rangle_{\leftarrow\text{ to }\rightarrow}^\mathrm{dip}
		\mp\,  
		\frac{J_\textit{NN}^\mathrm{dip}}{3} \, \hat{M}_{i,\perp} \, .
		\label{eq:point_dipole_barrier_torque}
	\end{equation}
For a central moment initially pointing to the left ($\leftarrow$), the second term (derived in Appendix~\ref{sec:derivation_chiral_splitting}) is subtracted for clockwise, and added for counter-clockwise reversal.

%
In conclusion, with a simplified point-dipole model the switching barriers are modified by the choice of clockwise vs.\ counter-clockwise rotation, if the moment interacts with an effective perpendicular stray field 
generated by its environment. 
As shown in Fig.~\ref{sfig:enumerate_of_states}, 40 out of the 64 double-vertex configurations feature a finite perpendicular magnetisation $\hat{M}_{i,\perp}\ne0$ acting on the central nanomagnet. 
In particular, we expect a maximum chiral barrier splitting for the fully-magnetised environments (marked in red), which are commonly-used initial states for thermal relaxation studies of artificial spin ice \cite{2013Farhan}.

These observations are by no means a curiosity, since nanomagnetic switching will occur predominantly via the more favourable pathway.
We thus expect profound consequences on the switching rates and transition kinetics when taking into account the chiral splitting.

\begin{figure*}[tb]
	\centering
	\includegraphics[width=174.781mm]{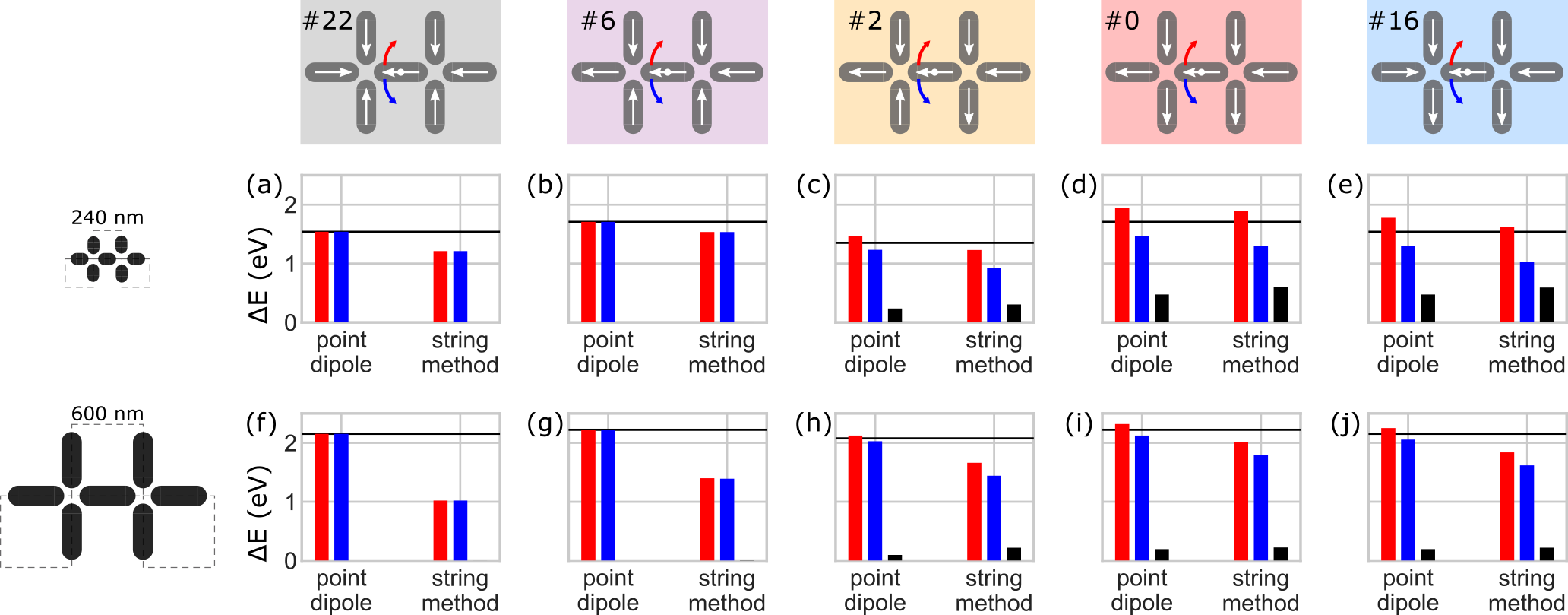}
	\caption{%
		\textbf{Comparison of switching barriers,} 
			for (a-e) exchange-dominated and (f-j) magnetostatic-dominated square ice and different environment states, as illustrated on top. The schematics on the left show the geometries drawn to scale.
			For each configuration the clockwise (red) and counter-clockwise barrier (blue) as well as their difference (black), obtained from point-dipole calculations (left) and string-method simulations (right), are plotted. 
			Switching barriers from the simplified point-dipole picture of uniform reversal systematically overestimate the micromagnetic simulation results, as well as underestimate the amount of barrier splitting for clockwise and counter-clockwise reversal (black bars).
			In general, the average point-dipole barrier, indicated with a black horizontal line, is an inadequate approximation to the switching barriers.
			}
	\label{fig:comparison_small_large_island}
\end{figure*}

\section{Micromagnetic switching}
\label{sec:simulations}

The point-dipole switching barriers represent a perturbative approach parametrised by two parameters only: First, $\Delta E_\text{sb}$ describes the switching barrier of an isolated nanomagnet, and implicitly depends on its shape and size \cite{1945Osborn}. Second, $J_\textit{NN}^\mathrm{dip}$ quantifies the energies of equilibrium configurations due to the interactions between nanomagnets placed on the square lattice, and modify the switching barrier of individual nanomagnets. 
The mean-field barrier energy in Eq.~(\ref{eq:point_dipole_barrier_torque}), however, does not take into account possible non-coherent moment reversal. 
In particular, it does not describe the influence of material parameters such as the saturation magnetisation $M_\text{sat}$ and the exchange strength $A_\text{ex}$, thermal fluctuations, and the magnetic environment. Due to these effects the net moment can be dynamically reduced during reversal, e.g.\ via non-uniform buckling modes, vortex creation, or domain formation \cite{1949Kittel,Hubert09,2009Guimaraes}. 

To have a nuanced look on how the barrier energy depends on (1)~the material parameters, (2)~the nanomagnet shape and size, and (3)~the interactions with neighbouring nanomagnets, we now turn to a full micromagnetic simulation of the reversal barriers.

\subsection{Implementation of string-method simulations}
\label{sec:string_method}


Contrary to simulations employing the Landau-Lifschitz-Gilbert equations to explicitly solve the time-dependent evolution of the nanomagnetic reversal, we determined the associated barrier energy using a time-independent string method. In this approach, starting from a coherent moment reversal, the moment configurations are iteratively optimised to a minimum-energy path through configuration space, and thus yield the lowest barrier energy associated to that reversal process. 
As in our previous work \cite{2020Koraltan}, which also discusses further simulation details, we implement here the \textit{simplified and improved string method (SISM)} \cite{2007Ren} using the finite-element micromagnetic code \texttt{magnum.fe} \cite{2013Abert}.

%
We consider two artificial square ice geometries with distinct choices for $M_\text{sat}$ and $A_\text{ex}$, representing different regimes. 
Meshes discretizing the considered geometries, i.e.\ an individual nanomagnet and the double-vertex configurations, were created with the software \texttt{gmsh} \cite{2009Geuzaine}.

First, we consider a geometry largely dominated by exchange interactions, which favour a coherent reversal, and thus may resemble the macrospin model derived in Sec.~\ref{sec:switching_theory}: 
Nanomagnets with dimensions $l\times w\times t=\SI{150x100x3}{nm}$ are placed on a square lattice with periodicity $a=\SI{240}{nm}$. The material parameters $M_\text{sat}=\SI{790}{kA/m}$ and $K=0$ correspond to bulk permalloy (Fe$_{0.2}$Ni$_{0.8}$) values at \SI{300}{K}. 
The exchange stiffness ${A_\text{ex}=\SI{13}{pJ/m}}$ was obtained from a temperature-dependent scaling \cite{1988Heider,2003Martinez,2016Moreno,2020Niitsu}
	\begin{equation}
	A_\text{ex} = A_\text{ex}(0) \left( \frac{M_\text{sat}}{M_\text{sat}(0)}  \right)^{1.7} \, ,
	\label{eq:Aex_scaling}
	\end{equation}
where $A_\text{ex}(0)=\SI{950}{kA/m}$ and $M_\text{sat}(0)=\SI{18}{pJ/m}$ denote the respective permalloy bulk parameters at \SI{0}{K}.

Second, we consider a system for which we expect sizable magnetostatic effects leading to non-uniform magnetic configurations during reversal:
Nanomagnets with dimensions $l\times w\times t=\SI{470x170x3}{nm}$ are placed on a square lattice with periodicity $a=\SI{600}{nm}$. 
This geometry, or choices close to it, have been used in several experimental studies \cite{2013Farhan,2017Gliga,2018Arava,2019Arava}.
The saturation magnetisation ${M_\text{sat}=\SI{350}{kA/m}}$ corresponds to the value given in \cite{2013Farhan}, and from the scaling in Eq.~(\ref{eq:Aex_scaling}) we obtain ${A_\text{ex}=\SI{3.25}{pJ/m}}$. Although the saturation magnetisation is lowered significantly, the exchange stiffness is even more reduced, thus we expect the energetics dominated by magnetostatic interactions.
We again assume a vanishing magnetocrystalline anisotropy, $K=0$.

\subsection{Influence of environment on reversal}
\label{sec:reversal_model_comparison}


We select representative environment states for each of the five classes introduced Fig.~\ref{sfig:enumerate_of_states}(b), for which we expect no (black and purple), intermediate (yellow), and high barrier splitting (red and blue), respectively. For these states, shown at the top of Fig.~\ref{fig:comparison_small_large_island}, we compare the chiral switching barriers obtained from a macrospin approximation (left) and micromagnetic string-method simulations (right) for the two different square ice geometries (schematics on the left are shown to scale). Clockwise and counter-clockwise barriers are marked  in red and blue, respectively, and the difference (barrier splitting) is plotted as black bars. 

The macrospin model considers variation of the single-moment barrier $\Delta E_\text{sb}$ (calculated from the shape anisotropy of a uniformly-magnetised nanomagnet ${\Delta E_\text{sb}^\text{shape}=K^\text{shape}V}$) due to point-dipole-like interactions quantified by ${J_\textit{NN}^\mathrm{dip}\propto(M_\text{sat}V)^2/a^3}$, as derived in Sec.~\ref{sec:point-dipole-barriers}. We also plot the mean-field barrier from Eq.~(\ref{eq:mean-field-barrier}) (black horizontal line).

The chiral barriers from the string-method simulations are calculated from the energy difference between the micromagnetic net energies of the metastable barrier configuration ($\uparrow$, $\downarrow$) and the initial configuration ($\leftarrow$):
\begin{eqnarray}
	\Delta E_{i, \leftarrow, \mathrm{cw}}^\mathrm{mm} 
	& = &  E_{i,\uparrow}^\mathrm{mm} - E_{i,\leftarrow}^\mathrm{mm} 	 \,	, 
	\label{eq:chiral_barriers_cw_mm}
	\\
	\Delta E_{i, \leftarrow, \mathrm{ccw}}^\mathrm{mm} 
	& = &  E_{i,\downarrow}^\mathrm{mm} - E_{i,\leftarrow}^\mathrm{mm} 	 \,	. 
	\label{eq:chiral_barriers_ccw_mm}
\end{eqnarray}


For the exchange-dominated square-ice geometry, Fig.~\ref{fig:comparison_small_large_island}(a-e), the mean-field barrier always overestimates the lower of the two micromagnetic barriers (for the chosen cases corresponding to counter-clockwise reversal marked in blue), as already discussed in our previous work \cite{2020Koraltan}. 
Compared to the point-dipole model, micromagnetic simulations consistently give lower chiral switching barriers. The difference of the barrier energies, i.e.\ the chiral splitting, is enhanced, however, it remains proportional to the perpendicular moment ${|\hat{M}_{i,\perp}|}$ generated by the environment.
The reversal process is still governed by an almost-coherent rotation of the central moment (Appendix~\ref{ssec:mphi}). Therefore, one can obtain reasonable switching barriers from the perturbative decomposition of Eq.~(\ref{eq:point_dipole_barrier_torque}) by using a lower single-moment barrier $\Delta E_\text{sb}^\text{string}$ and stronger interactions $J_\textit{NN}^\mathrm{mm}$ (Appendix~\ref{ssec:mm_modified_mean_field}). 


For the magnetostatic-dominated square-ice geometry, Fig.~\ref{fig:comparison_small_large_island}(f-j), the differences are even more pronounced.
This is because the reversal behaviour is no longer uniform, as discussed in  Appendices~\ref{ssec:mphi} and \ref{ssec:mm_modified_mean_field}. Interestingly, non-coherent reversal is particularly pronounced in environments with vanishing perpendicular magnetisation $\hat{M}_{i,\perp}=0$ (black and purple). This leads to large reductions of the switching barrier, e.g.\ more than \SI{-50}{\percent} in the case of state \#22 in Fig.~\ref{fig:comparison_small_large_island}(f), when compared to the point-dipole model.
In general, the point-dipole barriers overestimate all micromagnetic barriers by a considerate margin, and underestimate the chiral barrier splitting, as is evident when comparing the black bars shown in Fig.~\ref{fig:comparison_small_large_island}(h-j). For environments with a finite perpendicular magnetisation the barrier splitting is of similar magnitude, Fig.~\ref{fig:comparison_small_large_island}(h-j), and thus does not follow the proportionality ${|E_{i,\uparrow}-E_{i,\downarrow}|\propto|\hat{M}_{i,\perp}|}$. Therefore, the simplified decomposition of switching barriers of Eq.~(\ref{eq:point_dipole_barrier_torque}) is no longer valid.

In general, by taking into account the micromagnetic nature of the moment reversal, we observe both a reduction of the switching barriers as well an enhanced separation of the chiral barriers for environments with ${\hat{M}_{i,\perp}\ne 0}$. As lower barriers are easier to overcome for thermally-induced reversal, we therefore expect significant enhancement of the kinetics of artificial square ice, which may yield different relaxation time scales and emergent correlations.

\section{Transition kinetics}
\label{sec:kinetics}

The effect of barrier splitting on the net transition rate can be generalised by using an average barrier 
\begin{equation}
	\Delta E_\text{avg} = \frac{1}{2}
	\left| \Delta E_{i,\leftarrow,\mathrm{cw}} + \Delta E_{i,\leftarrow,\mathrm{ccw}} \right| \, ,
\end{equation}
and a factor $f$, which describes the symmetric splitting of the clockwise and counter-clockwise barriers around the average barrier, as depicted in Fig.~\ref{fig:rate_enhancement}(a),
\begin{eqnarray}
	f = 
	\frac{1}{2}\,
	\frac{ 
		\left| \Delta E_{i,\leftarrow,\mathrm{cw}} - \Delta E_{i,\leftarrow,\mathrm{ccw}} \right|
	}
	{
		\Delta E_\text{avg}
	} \, .
	\label{eq:rate_enhancement_factor}
\end{eqnarray}
For the barriers derived from the point-dipole picture in Sec.~\ref{sec:point-dipole-barriers}, the average barrier corresponds to the mean-field barrier of Eq.~(\ref{eq:mean-field-barrier}), i.e.\ ${\Delta E_\mathrm{avg}^\mathrm{dip}=\langle\Delta E_{i}\rangle_{\leftarrow\text{ to }\rightarrow}^\mathrm{dip}}$. The splitting factor vanishes, ${f=0}$, for environmental states without a perpendicular magnetisation ${\hat{M}_{i,\perp}=0}$ acting on the switching moment.
From the results of micromagnetic simulations discussed above we obtain non-zero values of $f$ between a few percent up to about ${\num{20}\%}$.

\subsection{Modified Arrhenius law for barrier splitting}
\label{sec:Arrhenius_enhancement}

\begin{figure}[t]
	\centering
	\includegraphics[width = 86mm]{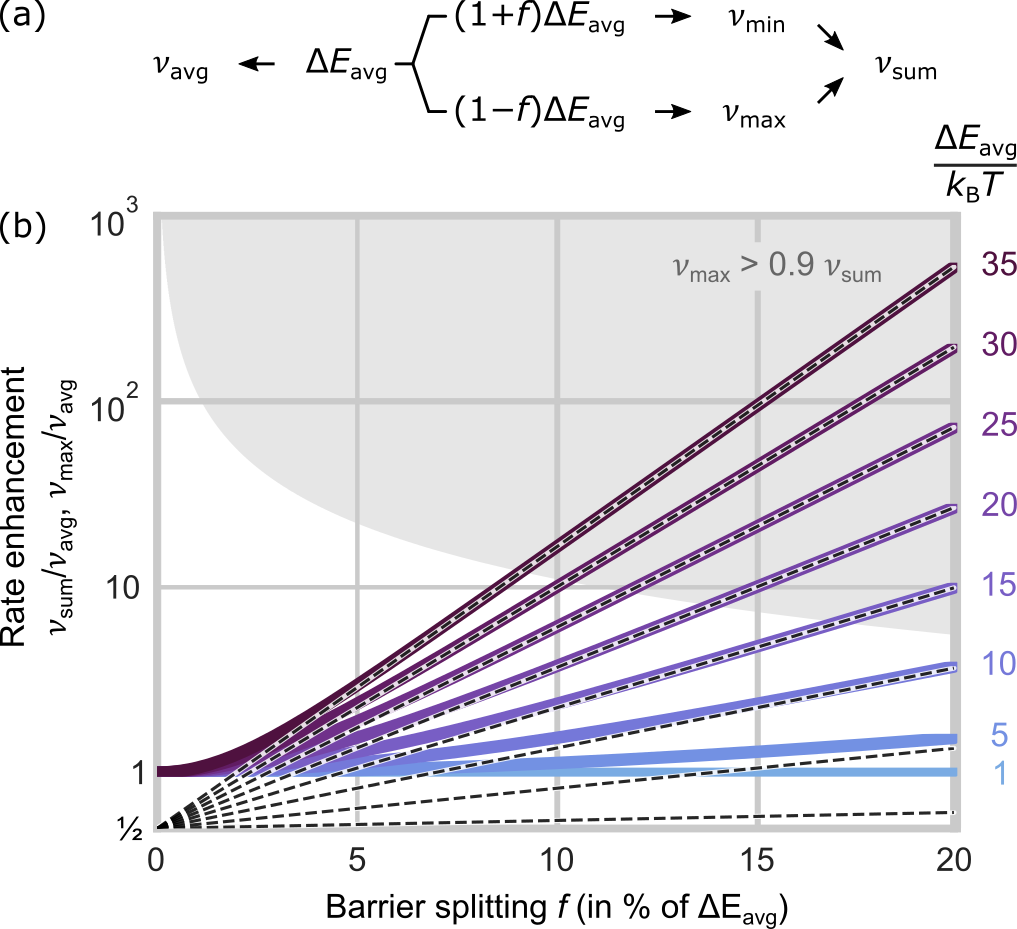}
	\caption{%
		\textbf{Rate enhancement due to barrier splitting.}
		(a)~Via the Arrhenius law, the transition rates $\nu_\text{avg}$, $\nu_\text{max}$ and $\nu_\text{min}$ can be derived for the different barrier energies.
		(b)~The sum rate $\nu_\text{sum}=\nu_\text{max}+\nu_\text{min}$ of the two parallel relaxation channels can be significantly enhanced compared to the rate expected from overcoming an average barrier $\nu_\text{avg}$, shown by the solid lines indicating $\nu_\text{sum}/\nu_\text{avg}$. 
		The colours denote different kinetic regimes given by the ratio of the average barrier $\Delta E_\mathrm{avg}$ compared to the thermal energy $k_\mathrm{B}T$, as indicated by the numbers on the right.
		Dashed lines denote the respective rate enhancement $\nu_\text{max}/\nu_\text{avg}$ associated with transitions via the \textit{lower} barrier only, which underestimates the rate by a factor of two in the limit of $f\rightarrow0$. Within the shaded area $\nu_\text{max}$ exceeds 90\% of the net rate $\nu_\text{sum}$.
		The rate enhancement is particularly large for rare events where the reduced energy, as indicated by the numbers on the right, is large, i.e.\ $\Delta E_\text{avg}/k_\mathrm{B}T\gg1$. However, it hardly matters in the limit of superparamagnetic fluctuations where $\Delta E_\text{avg}/k_\mathrm{B}T\rightarrow 1$.
	}
	\label{fig:rate_enhancement}
\end{figure}

\begin{figure*}[t!]
	\centering
	\includegraphics[width =175.471mm]{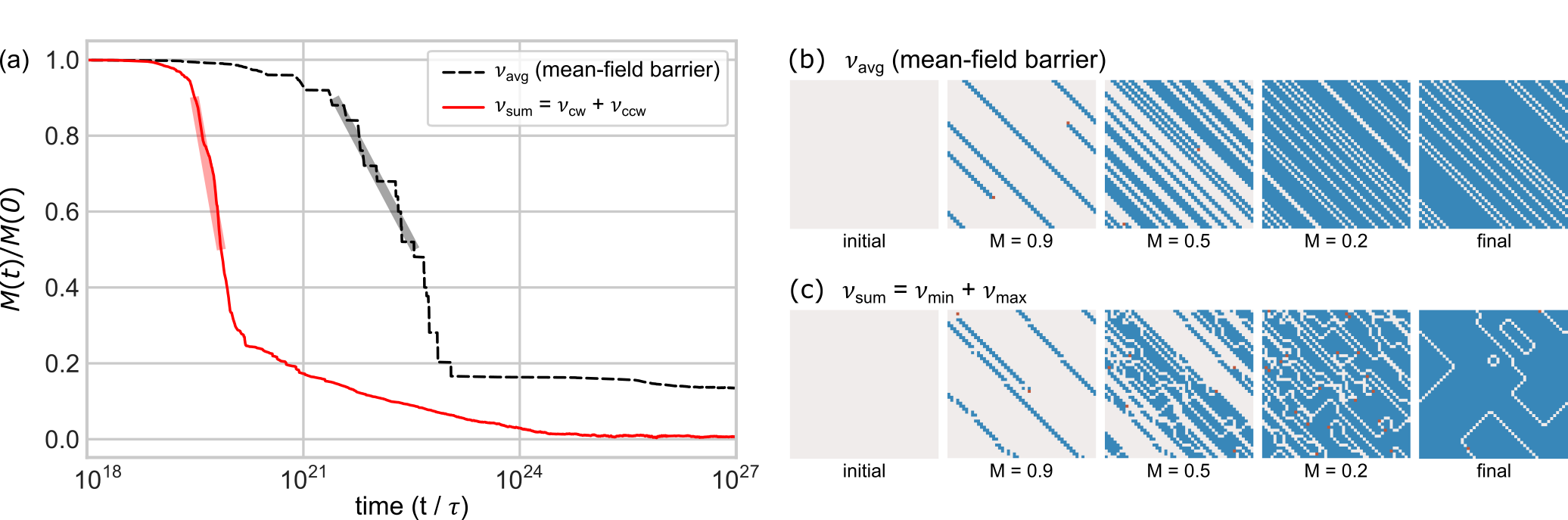}
	\caption{%
		\textbf{Evolution of extended square ice} from kinetic Monte Carlo simulations using point-dipole barriers calculated for the exchange-dominated geometry.
		(a)~Time-dependent net magnetisation, using switching rates obtained from the mean-field (dashed black line) and split-barrier model (solid red line). The time axis is normalised to the characteristic time scale given by the inverse attempt frequency $\tau=\nu_0^{-1}$. For the split-barrier model, the onset and rate of demagnetisation happens earlier and faster when compared to the mean-field-barrier model.
		(b,c)~Snapshots of spatial configurations for the (b)~mean-field and (c)~split-barrier model. Pixels correspond to 4-vertex spin arrangements. Those featuring a diagonal magnetisation or a ground-state configuration, are marked in gray or blue, respectively.
		}
	\label{fig:kMC-evolution}
\end{figure*}


The temperature-dependent transition rate for spontaneous switching over an average energy barrier $\Delta E_\mathrm{avg}$ can be obtained via the Arrhenius law \cite{2012Coffey,1963Brown}, with the attempt frequency $\nu_0$ and the Boltzmann constant ${k_{\mathrm{B}}=\SI{8.62e-5}{eV/K}}$:
\begin{eqnarray}
	\nu_\text{avg}(\Delta E_\text{avg}, T) 
		& = & 
		2 \nu_0 \exp\left( -\frac{\Delta E_\text{avg}}{k_\mathrm{B} T} \right) 
	\label{eq:Arrhenius_mean}
\end{eqnarray}
The attempt frequency $\nu_0$ depends on the shape, size, and material of the nanomagnets. Typical values of $\nu_0$ are in the order of $10^{9\ldots 12}\mathrm{~Hz}$ \cite{1963Brown,2018Stier}, and even faster time scales have been discussed \cite{2009Krause,2020Desplat}. 

%
We need to consider the clockwise and counter-clockwise reversal as parallel and independent channels of relaxation, and the rates associated to each of the two barrier energies, i.e.\ $\Delta E_\text{avg}(1-f)$ and $\Delta E_\text{avg}(1+f)$, need to be added to obtain an effective transition rate.
Therefore, for the definition of the rate in Eq.~(\ref{eq:Arrhenius_mean}), we explicitly included a pre-factor of two, to account for degenerate clockwise and counter-clockwise relaxation channels over the average barrier. 

Due to the pronounced non-linearity of the Arrhenius law, the summation of rates leads to an effective increase of the net transition rate $\nu_\mathrm{sum}$ of thermally-activated switching when compared to the rate $\nu_\text{avg}$ associated with the average barrier (derivation in Appendix~\ref{sec:derivation_arrhenius}):
\begin{equation}
	\nu_\mathrm{sum} 
	= \nu_\text{avg}(\Delta E_\text{avg}, T) \cosh\left(f\,\frac{\Delta E_\text{avg}}{k_\mathrm{B} T}\right) 
	\, .
	\label{eq:Arrhenius_split}
\end{equation}

The ratio of the joint rate compared to the average-barrier rate, i.e.\ $\nu_\mathrm{sum}/\nu_\mathrm{avg}$, are compared in Fig.~\ref{fig:rate_enhancement}(b) for different splitting ratios $f$ and reduced energies $\Delta E_\mathrm{avg}/k_\mathrm{B}T$.
The exponential rate enhancement is particularly pronounced in the limit of rare events with ${\Delta E_\mathrm{avg}/k_\mathrm{B}T\gg1}$ where a splitting of barriers can increase the (albeit low) transition rates by several orders of magnitude (purple lines).
In contrast, in the limit of superparamagnetic fluctuations, i.e.\ $\Delta E_\mathrm{avg}/k_\mathrm{B}T\rightarrow 1$, barrier splitting increases the net rate only moderately (blue lines).


For the assumption that transitions occur predominantly via the lower barrier \textit{only}, we have to consider the transition associated to the smaller barrier, i.e.\ $(1-f)\Delta E_\text{avg}$, see dashed lines in Fig.~\ref{fig:rate_enhancement} giving the ratio $\nu_\text{max}/\nu_\text{avg}$. 
For high splitting ratios $f$ and ${\Delta E_a/(k_B T)\gg1}$ the rate $\nu_\text{max}$ will approach $\nu_\text{sum}$, as transitions by the higher-lying barrier become irrelevant.
The shaded area of Fig.~\ref{fig:rate_enhancement} marks were $\nu_\text{max}$ exceeds 90\% of the value $\nu_\text{sum}$. Within this regime, transitions via the lower-lying barrier might be a good approximation of the net reversal rate.
In the limit of $f\rightarrow0$, however, where both barriers are equal, using the rate $\nu_\text{max}$ will underestimate the net rate by a factor of two, i.e. ${\left( \nu_\mathrm{max}/\nu_\mathrm{sum}\right)_{f\rightarrow0}=1/2}$.

In the case of artificial square ice, and depending on the kinetic regime given by the relation between the energies $\Delta E_\text{sb}$, $J_\textit{NN}$, and $k_\mathrm{B}T$, approximating the transition rates via the mean-field barrier \cite{2013Farhan,2013Farhan_a,2018Arava,2019Arava} or the minimum barrier \cite{2017Liashko,2018Gypens} thus may significantly underestimate the speed of evolution.

\subsection{Temporal evolution of extended square ice}

To illustrate the consequences of barrier splitting, we now turn to the evolution of extended square-ice arrays. 
In many thermal relaxation studies of artificial spin ice, the main interest lies in the onset of phase transitions and formation of emergent correlations.
In many experiments, the system evolves from a field-set fully magnetised state, for which we predict a particularly strong barrier splitting. We therefore expect that the initial demagnetisation of a magnetic-field-saturated artificial square ice array will be particularly affected by the modified transition kinetics.

To model the relaxation, kinetic Monte Carlo (kMC) simulations are often employed \cite{1975Bortz,1991Fichthorn,1999Newman}. The kMC algorithm provides a numerical solution to the master equation, which is a system of linear differential equations describing the evolution of the probabilities for Markov processes in systems that jump from one state to another in continuous time \cite{2014Toral}. Using this method, both the equilibrium expectation values of populations and their dynamical evolution during a thermalization process can be retrieved. 

In this work, kMC simulations are performed using a custom-written code \cite{Matteo-github}, with a system of \num{50x50} moments and periodic boundary conditions. The initial configuration is uniformly-magnetised, with the net magnetisation being parallel to a diagonal direction of the array. The demagnetisation due to spontaneous moment reversals is tracked over time for \num{125e3}~kMC steps, and averaged over \num{20} individual simulation runs.
We use the point-dipole energy barriers as shown in Fig.~\ref{sfig:point-dipole-splitting}(b) with parameters $J_\textit{NN}^\mathrm{dip}=\SI{0.178}{eV}$ and $\Delta E_\text{sb}=\SI{1.327}{eV}$ (i.e., using values for the exchange-dominated square-ice geometry), and calculate the environment-dependent transition rates $\nu_\text{avg}$ and $\nu_\text{sum}$ at a temperature of $T=\SI{300}{K}$ as input parameters for the kMC simulations.


Fig.~\ref{fig:kMC-evolution}(a) compares the time evolution of the net magnetisation of square ice for rates from the mean-field barriers (dashed black line) to the model taking into account chiral barrier splitting (solid red line). The time is measured in multiples of the inverse attempt frequency, $\tau=\nu_0^{-1}$.
We find that the onset of demagnetisation for the split-barrier model (red) happens two orders of magnitude earlier than for the average-barrier model (black). 
In the case of the average-barrier model, the demagnetisation involves bouts of rapid evolution interrupted by phases with little change, indicating avalanche-like dynamics \cite{2010Mengotti,2012Huegli,2014Chern,2014Budrikis}. In contrast, the split-barrier model shows a smooth demagnetisation, with a rate (solid thick lines indicate evolution from from 90\% to 50\%) which is about three orders of magnitude faster compared to the mean-barrier model.


When assessing the emergent spatial correlations, we find that the evolution for the mean-field model is governed by the propagation of a string of ground-state vertices (in blue) wrapping the system (due to the periodic boundary conditions), as shown in Fig.~\ref{fig:kMC-evolution}(b). The final state has a magnetisation of about \num{16}\% of its initial value. 
The snapshots of the spatial configuration of the split-barrier model at $M=\num{0.9}$ and $M=\num{0.5}$ (with $M$ normalised to the initial field-set magnetisation) in Fig.~\ref{fig:kMC-evolution}(c) appear somewhat similar to that of the mean-field case. There are more possible transitions for the system to explore, however, and the final state of the evolution corresponds to a multi-domain state with almost vanishing magnetisation, $M\approx0$.


Thus, our kMC results show that the modified hierarchy of transition barriers due to the chiral barrier splitting may have subtle, but relevant, consequences: In certain cases, the kinetic relaxation pathways are not simply dictated by equilibrium-energy arguments.
%
This will modify the emergence of spatial correlations, which needs to be explored in a systematic study and compared to experimental results \cite{2006Wang,2010Morgan,2010Mengotti,2012Budrikis,2012Kapaklis,2012Nisoli,2014Kapaklis,2016Vedmedenko,2016Andersson,2019Zhang,2020Martinez,2020Arava}.

\section{Conclusions} 
\label{sec:conclusions}


To realistically model the temporal evolution of artificial spin ices or small-scale nanomagnetic circuits it is necessary to know the switching barriers for the single-moment reversal.
In this work, we quantified how magnetostatic interactions with neighbouring nanomagnets modify the switching barriers in artificial square ice in absence of \textit{extrinsic} effects such as defects or spurious fields.
%
%
We found that for environments which feature a finite perpendicular magnetic field acting on the switching nanomagnet clockwise and counter-clockwise moment reversals need to be considered independently. The resulting barrier splitting can be sizeable. In the case of exchange-dominated nanomagnets supporting coherent rotation modes the splitting can be predicted from a modified point-dipole model.
Taking into account the finite size of the nanomagnets and the influence of material parameters, further barrier reductions were obtained from micromagnetic simulations. These reductions are particularly strong for magnetostatically-dominated nanomagnets embedded in environments that do not promote reversal via a distinct chiral switching channel.


The splitting and reduction of transition barriers exponentially increase the transition rates when compared to a mean-field average barrier. Depending on the dynamical regime, which depends on the relationship between the average barrier energy, barrier splitting and temperature, we found that transition rates are especially enhanced in the limit of rare events.
We modelled the evolution of extended artificial square ice with kinetic Monte Carlo simulations, and compared a mean-field model with the model that takes into account the barrier splitting. We found that the onset and speed of evolution is largely enhanced in the latter case. 
Furthermore, while mean-field barriers are solely dictated by equilibrium-energy arguments, the chiral switching barriers depend on the kinetics of reversal. Thus, more and different relaxation pathways are accessible, which modifies the emergent spatial correlations and routes towards the ground state.


Our results are a step towards a deeper understanding of the single-moment switching of nanomagnetic systems, highlighting how faster time scales of relaxation can be caused via \textit{intrinsic} interactions with the magnetic environment.
These findings are relevant to the field of artificial spin systems, and can be extended from artificial square ice to other moment configurations, such as kagome ice \cite{2010Mengotti,2012Huegli,2013Zhang,2017Farhan}, and square-ice-like tetris, shakti, and brickwork lattices featuring asymmetric moment coordinations \cite{2014Gilbert,2015Gilbert_a,2018Lao}.
We also expect that these concepts are relevant for the utilisation of magnetic metamaterials for magnonics \cite{2009Neusser,2010Kruglyak,2013Gliga,2016Haldar,2016Jungfleisch,2019Lendinez,2020Iacocca} and nanomagnetic computation \cite{2006Imre,2018Arava,2018Gypens,2019Arava,2019Pancaldi}.

\begin{acknowledgments}

M.P.\ gratefully acknowledges David De Sancho for the assistance during the development of the kinetic Monte Carlo simulation code.
The computational results presented have been in part achieved using the Vienna Scientific Cluster (VSC).
N.L.\ has received funding from the European Union’s Horizon 2020 research and innovation programme under the Marie Słodowska Curie Grant Agreement No.~844304 (LICONAMCO).
N.L., M.M., and P.V.\ acknowledge support from the Spanish Ministry of Economy, Industry and Competitiveness under the Maria de Maeztu Units of Excellence Programme (MDM-2016-0618), and the Spanish Ministry of Science and Innovation funding the pre-doctoral grant PRE2019-088070 and the project RTI2018-094881-B-I00 (MINECO/FEDER).
%
%
K.H. acknowledges funding by the Swiss National Science Foundation (Project No.\ 200020\_172774).

\end{acknowledgments}

%
%
%
%

\appendix

\begin{figure}[tb]
	\centering
	\includegraphics[width=77.867mm]{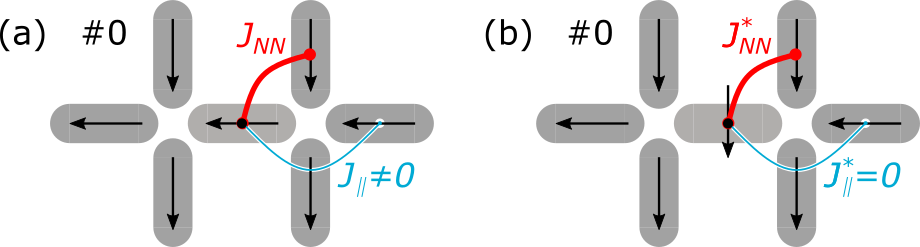}
	\caption{ %
		\textbf{Definition of interactions $J_\textit{NN}$ and $J_\textit{NN}^\ast$.}
		Nearest-neighbour dipolar interactions for (a)~the equilibrium and (b)~the high-energy configuration. 
		In the latter, the staggered arrangement of moments gives rise to a preferred ferromagnetic head-to-tail arrangement in the excited configuration, with a modified dipolar nearest-neighbour interaction strength ${J_\textit{NN}^\ast=J_\textit{NN}/3}$. 
		Interactions of the central moment with horizontal moments via ${J^\ast_\parallel=0}$ vanishes.
		}
	\label{sfig:JNN-JNNstar}
\end{figure}

\begin{figure*}[t]
	\centering
	\includegraphics[width=178mm]{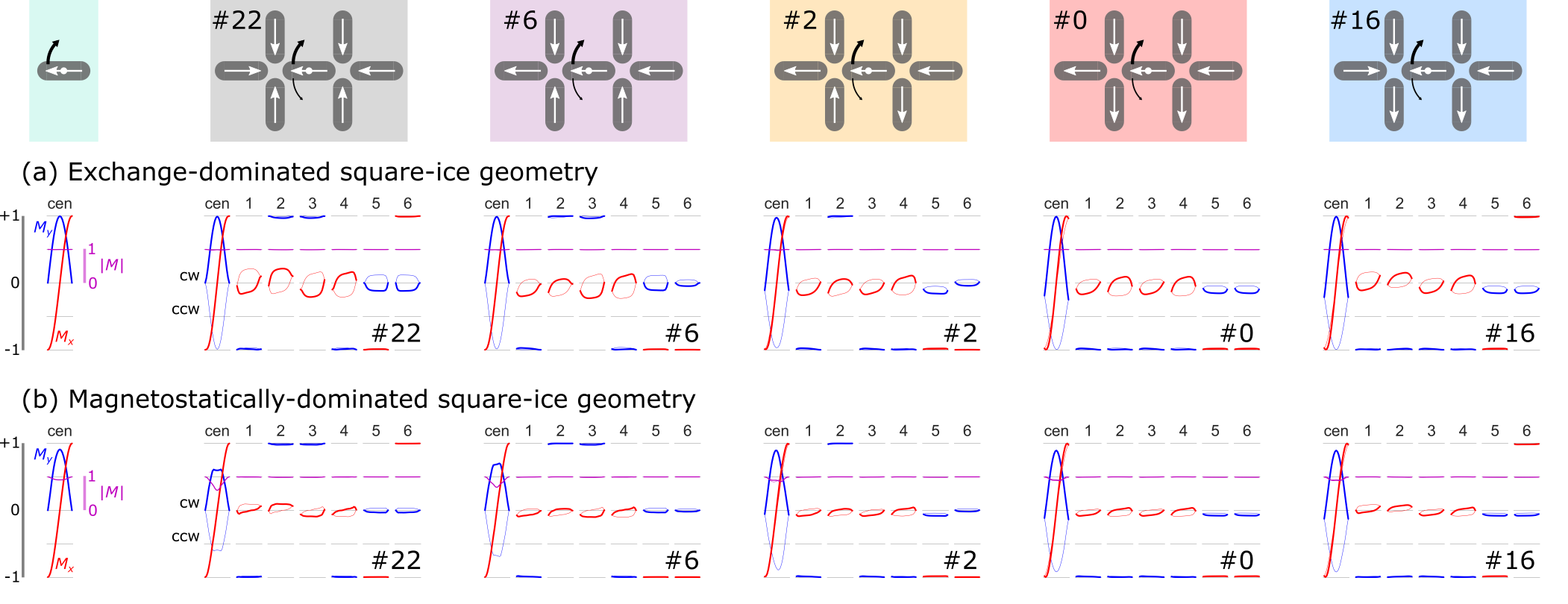}
	\caption{%
		\textbf{Variation in magnetisation during moment reversal} for (a)~exchange- and (b)~magnetostatic-dominated square-ice geometries.
		Plotted are the average magnetisation of an isolated nanomagnet (left) and of the individual nanomagnets obtained from the minimum-energy path simulations for each the representative configurations in Fig.~\ref{fig:comparison_small_large_island}.
		The average moment $M_x$ along the horizontal (i.e.\ parallel to the easy axis of the central nanomagnet) and $M_y$ along the vertical direction are plotted in red and blue, respectively. Values for clockwise (counter-clockwise) rotation from left to right are plotted by thick (thin) lines.
		The average net moment $|M|$ of each nanomagnet is indicated by magenta lines (note the reduced scale annotated on the left of the isolated nanomagnet).
	}
	\label{sfig:mphi}
\end{figure*}

\begin{figure*}[t]
	\centering
	\includegraphics[width=173.219mm]{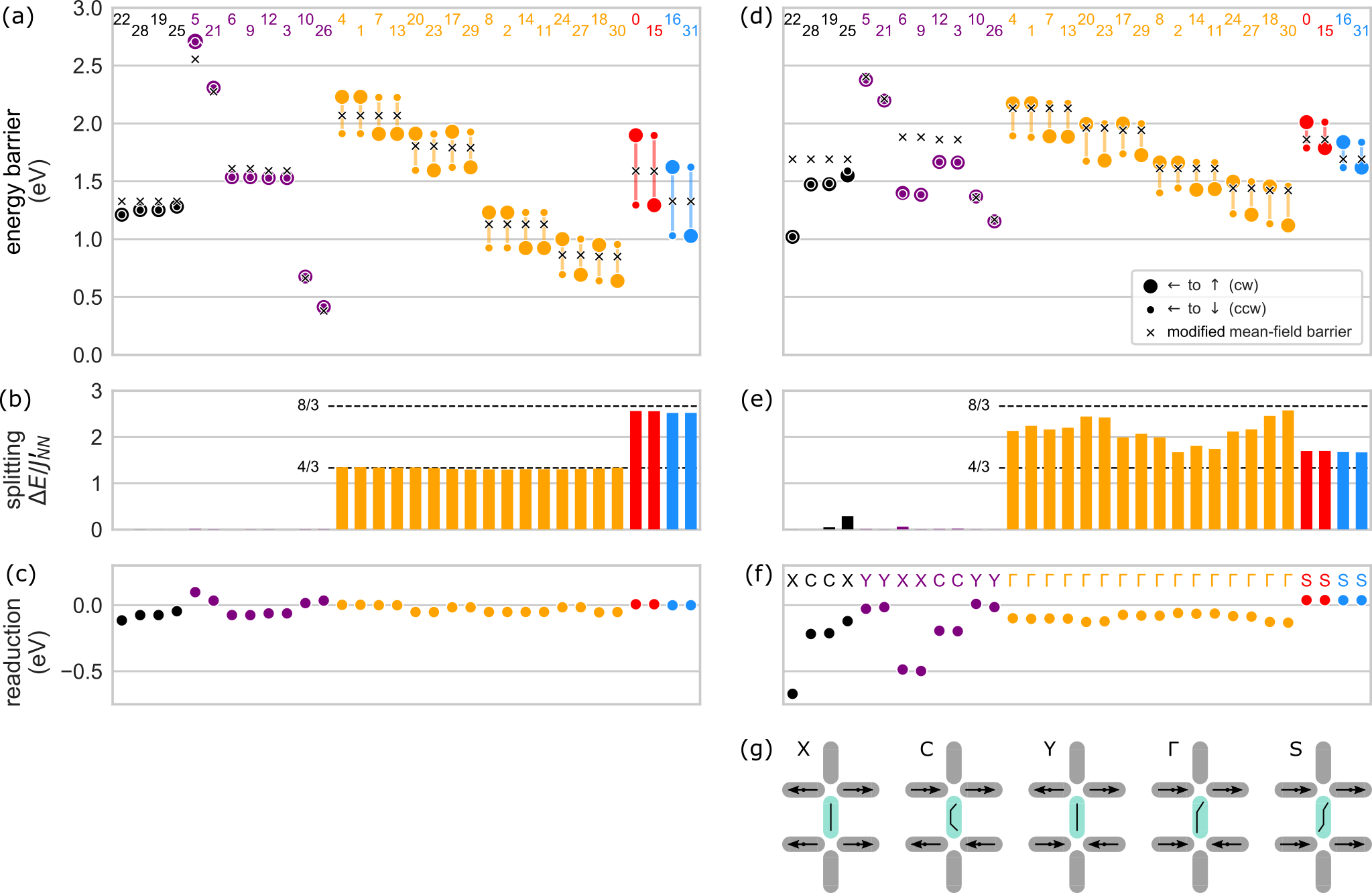}
	\caption{ %
		\textbf{Switching barriers from micromagnetic string-method simulations,} for (a-c)~exchange- and (d-f)~magnetostatic-dominated square-ice geometries.
		(a,d)~Barrier energies for the moment reversal from left to right via clockwise and counter-clockwise pathways (large and small circles, respectively). The modified mean-field barrier, discussed in Appendix~\ref{ssec:mm_modified_mean_field}, is marked with a cross.
		(b,e)~Difference of the two barriers, compared to the modified mean-field prediction of the barrier splitting (dashed lines).
		(c,f)~Difference between the modified mean-field barrier [marked by crosses in (a,d)] and the average micromagnetic barrier of clockwise and counter-clockwise reversal [i.e., centre of vertical lines in (a,d)]. Large differences are observed especially for those environments which do not promote transitions of preferred chirality (marked in black and purple).
		(g)~Nomenclature of the environment states used in (f), determined by the relative orientation of the perpendicular nanomagnets.
	}
	\label{sfig:micromagnetics_all}
\end{figure*}

\section{Derivation of point-dipole barriers}


We assume that all moments are strictly parallel ($\leftarrow$, $\rightarrow$) to the nanomagnet long axis, and remain static during the reversal of the central moment. We furthermore assume that the configuration of highest energy correspond to those with perpendicular central moment ($\uparrow$, $\downarrow$). 
These approximations are approximately valid for weak interactions only. In general, however, they are a gross oversimplification, as due to the pairwise couplings the macrospins may rotate away from the local symmetry axis. This would result e.g.\ non-symmetric splitting for clockwise and counter-clockwise transitions (i.e.\ $\Delta \phi\ne\pi$).
Nevertheless, the strict limitation of moment direction allows to employ the anti-symmetry of the dipolar interaction energy under moment rotations of $\pi$, 
\begin{equation}
	E^\mathrm{dip}(\phi + \pi) = -E^\mathrm{dip}(\phi) \, .
	\label{seq:symmetry_argument}
\end{equation}

\label{sec:derviation_mean_barrier}

The mean-field barrier ${\langle\Delta E_{i}\rangle_{\leftarrow\text{ to }\rightarrow}^\mathrm{dip}}$ is the average of clockwise and counter-clockwise energy barriers. Under the above assumptions, and as derived in Appendix~B of Ref.~\cite{2020Koraltan}, it is determined by $\Delta E_\mathrm{sb}$ and the energy difference between the equilibrium states before and after switching. It does not depend, however, on the energies of the intermediate high-energy configuration.

\label{sec:derivation_chiral_splitting}

The barrier splitting $\Delta E_\mathrm{split}^\mathrm{dip}$ between clockwise and counter-clockwise rotation, see Eq.~(\ref{eq:point_dipole_barrier_torque}), can be calculated from the energy difference of the high-energy states. Using the anti-symmetry argument of Eq.~(\ref{seq:symmetry_argument}), i.e.\ ${E_{i,\uparrow}^\mathrm{dip}=-E_{i,\downarrow}^\mathrm{dip}}$, one obtains
\begin{equation}
	\Delta E_\mathrm{split}^\mathrm{dip} 
	=
	\left| E_{i,\uparrow}^\mathrm{dip} - E_{i,\downarrow}^\mathrm{dip} \right| 
	=
	2 \, \left| E_{i,\uparrow}^\mathrm{dip,*} \right| \, .
	\label{eq:split_deriv_1} 
\end{equation}
Here, $E_{i,\uparrow}^\mathrm{dip,*}$ takes into account the dipolar interaction terms with the central moment only (couplings between other moments remain unchanged by the reversal, and thus fall out of the energy difference).

As shown in Fig.~\ref{sfig:JNN-JNNstar}(b), in the high-energy state interactions with the horizontal nanomagnets, i.e.\ moments $5$ and $6$ in Fig.~\ref{fig:square_ice-overview}(b), will vanish as $J_\parallel^*=0$.
Due to the staggered arrangement of the moments, a ferromagnetic head-to-tail alignment of the central nanomagnet to its perpendicular neighbours is favourable, whereas the opposite orientation is penalised, see Fig.~\ref{fig:square_ice-overview}(c).
Thus, the net perpendicular magnetisation ${\hat{M}_{i,\perp}=\Sigma_{j=1}^{4}(-1)^{b_j}}$ is relevant to the splitting only. With a modified pair-wise interaction ${J_\textit{NN}^\mathrm{dip,*} = J_\textit{NN}^\mathrm{dip}/3}$ one obtains
\begin{eqnarray}
	\Delta E_\mathrm{split}^\mathrm{dip}
	& = & 
	\frac{2}{3} J_\textit{NN}^\mathrm{dip} \, \hat{M}_{i,\perp} \, .
	\label{eq:split_deriv_3}
\end{eqnarray}

\section{Arrhenius law for barrier splitting}
\label{sec:derivation_arrhenius}

If the transition barriers split symmetrically by a fraction $f$ around the average barrier ${\Delta E_\text{avg}}$ to values ${\Delta E_\mathrm{cw/ccw} =  (1 \pm f) \Delta E_\text{avg}}$, the joint effective rate ${\nu_\text{sum} = \nu_\mathrm{cw} + \nu_\mathrm{ccw}}$ can be expressed as follows:
	\begin{eqnarray}
	\nu_\text{sum} 
	& = & \nu_\mathrm{cw} + \nu_\mathrm{ccw} \\
	& = & \nu_0 \left[ e^{-(1+f)C} + e^{-(1-f)C} \right] \\
	& = & \nu_0\, e^{-C} \, \left( e^{-fC} + e^{+fC} \right) \\
	& = & 2 \nu_0 \, e^{-C} \, \cosh(fC)\\
	\nu_\text{sum} & = &  
	\nu_\text{avg}(\Delta E_\text{avg}, T) \cosh\left(f\, \frac{\Delta E_\text{avg}}{k_\mathrm{B} T}\right) 
	\end{eqnarray}
Here, $C = \Delta E_\text{avg} / (k_\mathrm{B} T)$ denotes the reduced average switching barrier energy. 
We assume that the attempt frequencies $\nu_0$ are independent of the energy of the saddle point, i.e.\ $\nu_0^\mathrm{cw}=\nu_0^\mathrm{ccw}=\nu_0$.
The transition rate $\nu_\text{avg}(\Delta E_\text{avg}, T)$ is defined in Eq.~(\ref{eq:Arrhenius_mean}).

The maximum of the clockwise and counter-clockwise switching rates is associated to the lower-lying barrier energy ${(1-f)\Delta E_\text{avg}}$. 
In the limit of $f\rightarrow 0$, $\nu_\mathrm{max}$ is a factor of two smaller than the rate ${\nu_\text{sum}(f=0)}$, and approaches the value of $\nu_\text{sum}$ for large splitting $f$ or large reduced energy ${C\gg 1}$:
	\begin{eqnarray}
	\nu_\text{max} 	& = & \max(\nu_\text{cw},\nu_\text{ccw}) \\
					& = & \nu_0 e^{-C(1-f)} = \nu_0 e^{-C} e^{fC} \\
					& = & \nu_\text{sum} \left( 1 +  e^{-2fC} \right)^{-1} \\
					& = & 
					\begin{cases}
						\frac{1}{2}\nu_\text{sum} & \text{for} \;  f\rightarrow 0\\
						\nu_\text{sum} & \text{for} \; fC\gg 1 					
					\end{cases}
					\; .
	\end{eqnarray}

\section{Additional simulation results}
\label{sec:mm_details}


\subsection{Magnetisation during reversal}
\label{ssec:mphi}


To quantify the uniformity of the magnetic reversal, Fig.~\ref{sfig:mphi} shows the averaged moments for a single (non-interacting) nanomagnet and each of the representative double-vertex configurations presented in Fig.~\ref{fig:comparison_small_large_island}.
Here, the average magnetisation of each nanomagnet is plotted for every step of the string-method minimum-energy path. The horizontal coordinate roughly corresponds to the rotation angle $\phi$ of the central moment, with end points denoting initial and final equilibrium states. 

In general, we find that in equilibrium the net moment ${|\vec{M}|}$ (magenta lines) is very close to one. Therefore, the static nanomagnets assume an almost saturated configuration, with the magnetisation largely aligned with the long axis of the nanomagnet and limited edge bending ($M_x$, red lines). Interactions with neighbouring moments, however, can induce sizeable perpendicular moment contributions (${M_x\perp M_y}$, blue lines) in environments that feature a finite perpendicular magnetisation ${\hat{M}_{i,\perp}\ne0}$, i.e.\ configurations \#0, \#2, and \#16.


In the case of the geometry with small islands dominated by exchange interactions the magnitude $|M|$ remains largely constant, Fig.~\ref{sfig:mphi}(a). The reversal thus represents a quasi-uniform rotation of the central moment. During reversal, the magnetisation components of the neighbouring nanomagnet can vary, allowing the system to evolve via the most efficient pathway. 


For the geometry with large islands dominated by magnetostatic energy, shown in Fig.~\ref{sfig:mphi}(b), the switching of the non-interacting nanomagnet (left) involves a reduction of the net moment to ~91\%, and thus does not conform to a uniform moment rotation. 
For environment states \#0, \#16, and \#2 with ${\hat{M}_{i,\perp}\ne0}$ the reduction of net magnetisation is similar to that of an individual nanomagnet. In contrast, for environment states \#6 and \#22, with  ${\hat{M}_{i,\perp}=0}$, we observe a pronounced reduction of magnetisation in the high-energy configuration to less than 70\% of the net moment, indicating non-coherent reversal. This leads to a reduction of the switching barrier as well, as discussed in Sec.~\ref{sec:simulations}.
The magnetic configuration of neighbouring nanomagnets varies less during reversal when compared to the small-island geometry. This is because the relative volume fraction of the large magnets meeting at the vertex point, where the spin structure varies the most, is smaller.

\subsection{Micromagnetic barriers}
\label{ssec:mm_modified_mean_field}


The switching barriers obtained from micromagnetic string-method simulations for the environment states \#0-\#31 are summarised in Fig.~\ref{sfig:micromagnetics_all} with (a-c) showing the results for the exchange-dominated, and (d-f) the magnetostatic-dominated geometry.

We compare the clockwise and counter-clockwise barriers, large and small circles in Fig.~\ref{sfig:micromagnetics_all}(a,d), to a \textit{modified} mean-field model. The predictions are based on Eqs.~(\ref{eq:mean-field-barrier}) and (\ref{eq:point_dipole_barrier_torque}), but instead of energies derived from point-dipole calculations we use those obtained from micromagnetic simulations, as follows:

First, the switching barrier $\Delta E_\text{sb}$ of an isolated nanomagnet simulated with the string-method is used, as opposed to the shape anisotropy calculated for a uniformly-magnetised nanomagnet. 
For the exchange-dominated nanomagnet geometry we obtain ${\Delta E_\text{sb}^\text{shape}=\SI{1.540}{eV}}$ compared to ${\Delta E_\text{sb}^\text{string}=\SI{1.327}{eV}}$, which is a reduction of \num{-14}\%. 
For the magnetostatic-dominated nanomagnet geometry, the barrier reduction is even bigger, with ${\Delta E_\text{sb}^\text{shape}=\SI{2.153}{eV}}$ and ${\Delta E_\text{sb}^\text{string}=\SI{1.691}{eV}}$ (\num{-22}\%), as the reversal is no longer coherent [see Fig.~\ref{sfig:mphi}(b)].

Second, $E_{i,\leftarrow}^\mathrm{mm}$ and $E_{i,\rightarrow}^\mathrm{mm}$ correspond to the micromagnetic equilibrium energies of the static configurations. Together with $\Delta E_\mathrm{sb}^\text{string}$, a modified mean-field barrier can be calculated from Eq.~(\ref{eq:mean-field-barrier}) [crosses in Figs.~\ref{sfig:micromagnetics_all}(a,d)].

Third, to estimate the barrier splitting, the nearest-neighbour interaction $J_\textit{NN}^\mathrm{mm}$ is rescaled. The re-scaling is motivated by the point-dipole model, which predicts an energy difference of ${( 16 - \frac{24 \sqrt{5}}{125} )J_\mathit{NN}^\text{dip}}$ between the lowest-lying ground state and highest monopole state: 
\begin{equation}
J_\textit{NN}^\mathrm{mm} = \frac{ E_\text{max}^\mathrm{mm} - E_\text{min}^\mathrm{mm} }{  16 - \frac{24 \sqrt{5}}{125} } \, .
\label{seq:JNN_dv}
\end{equation}

For the exchange-dominated small-island geometry we find that the modified mean-field barrier gives a passable estimate for the average switching barrier, as the small differences in Fig.~\ref{sfig:micromagnetics_all}(c) show. The chiral barrier splitting is well-described by the re-scaled energy $J_\textit{NN}^\mathrm{mm}$, albeit with a small reduction for fully-magnetised environments marked red and blue in Fig.~\ref{sfig:micromagnetics_all}(c).

For the magnetostatic-dominated reversal in the large-island geometry, Fig.~\ref{sfig:micromagnetics_all}(d-f), the mean-field approach fails: The barrier splitting is both overestimated for environments with $|\hat{M}_{i,\perp}|=4$ (red and blue) as well as underestimated in the case of $|\hat{M}_{i,\perp}|=2$ (yellow). In particular, the mean-field barrier predictions fails for environments with $|\hat{M}_{i,\perp}|=0$ (black and purple), where reversal via non-uniform modes are favoured, as discussed before.
The reductions compared to the mean-field barrier seem to be particularly strong for environments which feature "X" and "C" configurations of the perpendicular moments, see Figs.~\ref{sfig:micromagnetics_all}(f,g). This highlights the importance of considering the magnetostatic interactions with neighbouring nanomagnets during reversal to obtain the correct barrier energies. 

With the exception of $\Delta E_\text{sb}^\text{string}$, which is a result of the string-method simulation, the energies $E_{i,\leftrightarrow}^\mathrm{mm}$ and $J_\textit{NN}^\mathrm{mm}$ can be obtained from static equilibrium micromagnetic simulations, e.g.\ using OOMMF \cite{oommf_userguide} or MuMax3 \cite{2014Vansteenkiste}. 
This makes this approach attractive to estimate more realistic switching barriers based on a pertubative decomposition of a single-nanomagnet behaviour plus a correction term due to interactions with the neighbouring moments. 
This approach seems valid for relatively small nanomagnets favouring reversal via uniform modes, but fails if more complex reversal mechanisms are accessible.


%

\end{document}